\documentstyle[epsf]{aipproc}

\begin{document}
\title{Cosmological Matter Perturbations with Causal Seeds}
\author{Jiun-Huei Proty Wu}
\address{Astronomy Department,
  University of California, Berkeley,
  CA 94720-3411, USA}

\maketitle

\begin{abstract}
We investigate linear matter density perturbations
in models of structure formation with causal seeds.
Under the fluid approximation,
we obtain the analytic solutions using Green-function technique.
Some incorrect solutions in the literature are corrected here.
Based on this,
  we analytically prove that
  the matter density perturbations today are independent of the way 
  the causal seeds were compensated into the background contents
  of the universe when they were first formed.
We also find that
  the compensation scale depends 
not only on the dynamics of the universe,
but also on the properties of the seeds near the horizon scale.
It can be accurately located by employing our Green-function solutions.
\end{abstract}

With the cosmological principle as the basic premise,
there are currently two main paradigms for 
structure formation of the universe---inflation \cite{Guth}
and topological defects \cite{VilShe}.
Although recent observations of the cosmic microwave background (CMB)
seem to have favored inflation \cite{maxiboom},
defects can still coexist with it 
and their observational verification will have certain impact to
 the grand unified theory.
In the literature
the power spectra of models with causal seeds like defects
have been investigated using the full Einstein-Boltzmann equations.
However,
the study of the phase information of these perturbations 
still remains difficult due to the limited computation power \cite{AveShe4}.
Although
there have been some detailed treatments for models with causal seeds
\cite{HuSelWhi,HuWhi},
we shall present a simpler formalism
to provide not only a physically transparent way 
to understand the evolution of their density perturbations,
but also a computationally economical scheme
to investigate their phase information \cite{cmp}. 
This formalism is parallel to
those presented in \cite{VeeSte} and \cite{turok},
but 
we give modifications
to include the cosmological constant $\Lambda$,
as well as some other improvements and corrections.

\begin{figure}[t]
  \centering
  \leavevmode\epsfxsize=2.9in\epsfbox{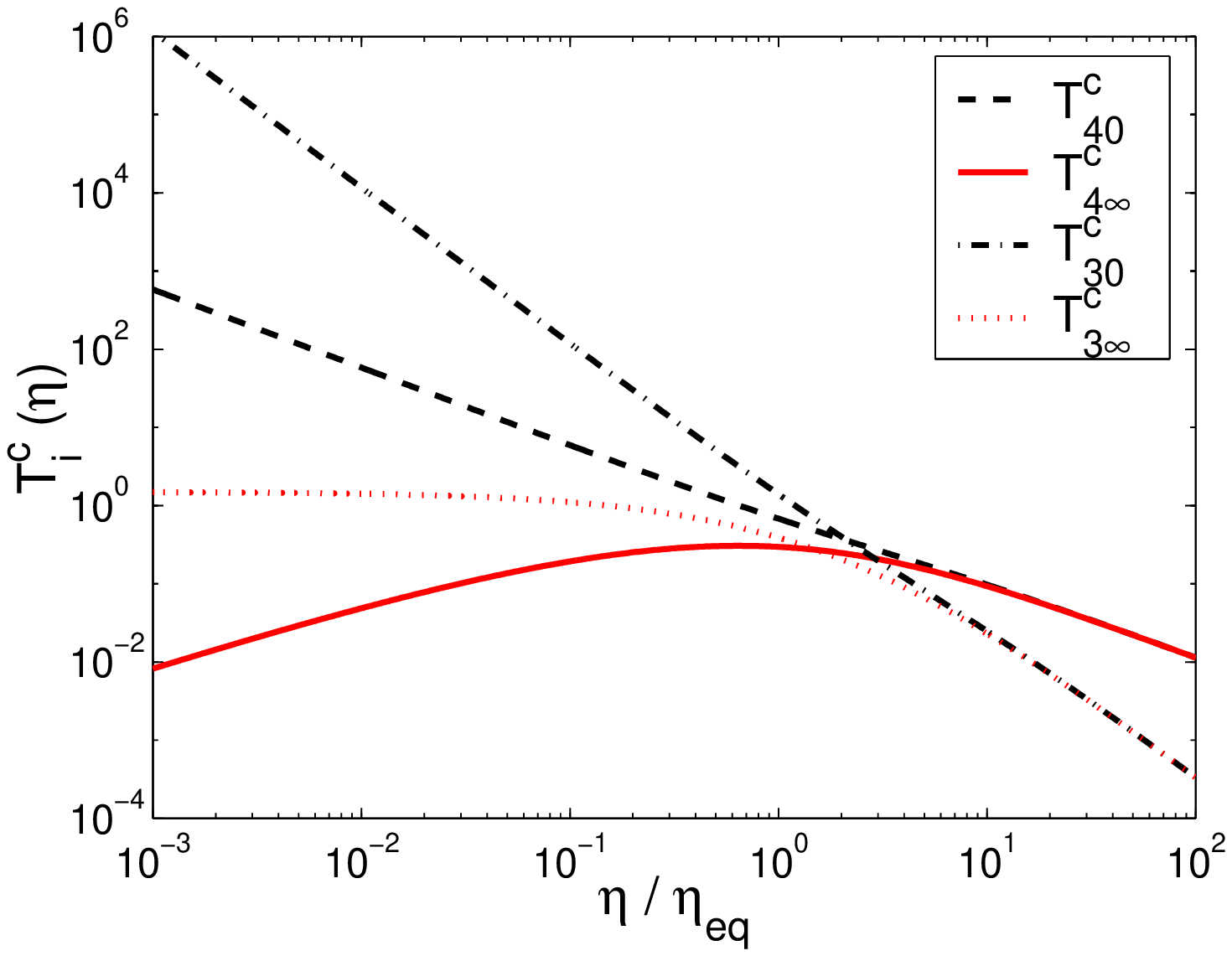}\hspace*{2mm}
  \leavevmode\epsfxsize=2.6in\epsfbox{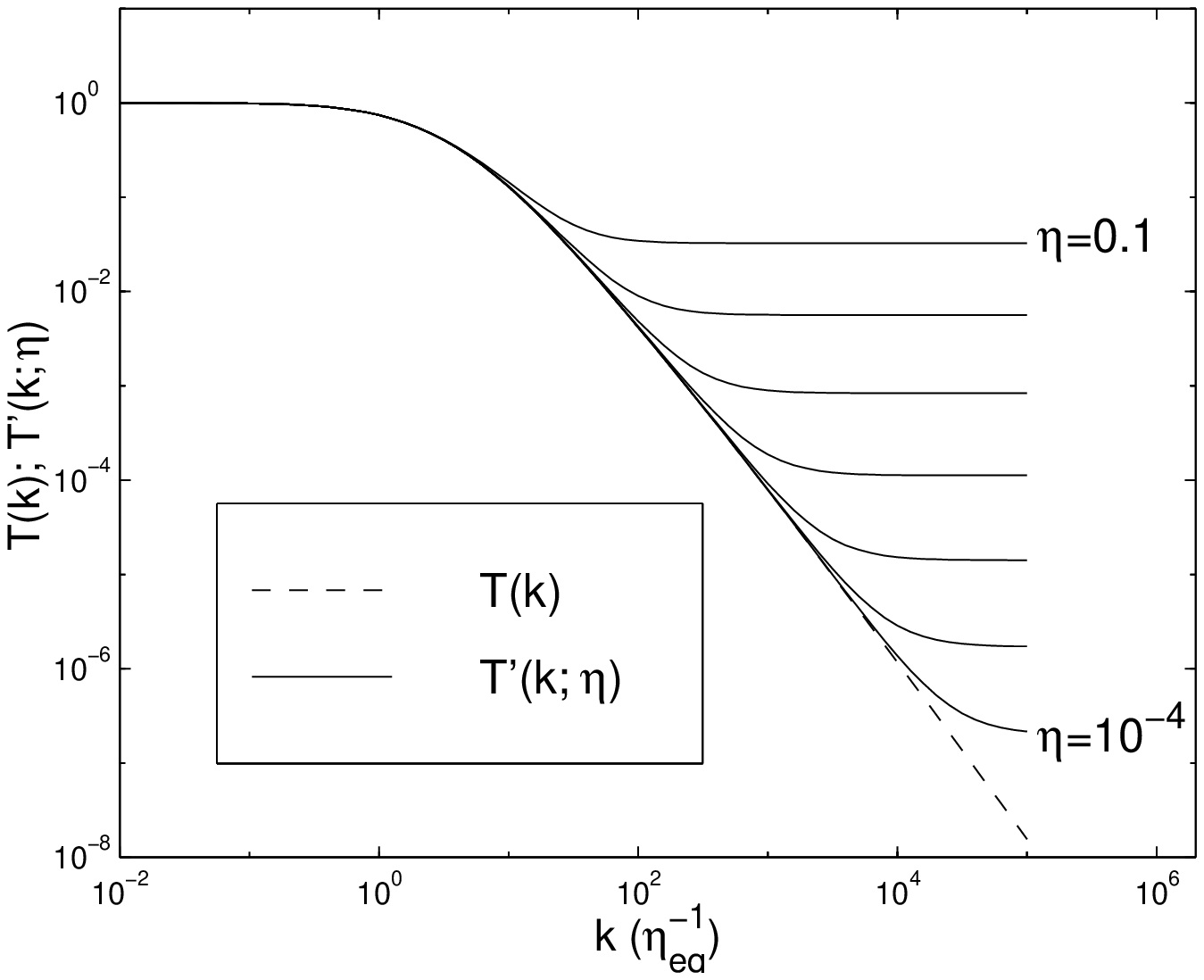}\\
  \caption
  {The asymptotic behaviors of the CDM Green-function solutions
   on the super-horizon and sub-horizon scales (left),
   as well as on all scales (right).
  }
  \label{fig-1}
\end{figure}

In a flat Friedmann-Robertson-Walker (FRW) model 
with an evolving weak source field of 
energy-momentum tensor $\Theta_{\mu\nu}({\bf x}, \eta)$,
the full evolution equations of linear perturbations
can be obtained by
considering the stress-energy conservation of 
the fluids and the source,
as well as the linearly perturbed Einstein equations \cite{cmp}.
With the photon-baryon tight coupling approximation,
a closed set of equations for the density perturbations 
in the synchronous gauge are:
\begin{eqnarray}
  \ddot \delta_{\rm r} - {4 \over 3} \ddot \delta_{\rm c}
  +\frac{\dot R}{1+R}(\dot \delta_{\rm r} - {4 \over 3} \dot \delta_{\rm c})
  - {1 \over 3(1+R)} \nabla^2 \delta_{\rm r}
  =  0,
  \label{delta-one}\\
  \ddot \delta_{\rm c}
  + {\dot a \over a} \dot \delta_{\rm c} 
  - {3 \over 2}\Big({\dot a \over a}\Big)^2 
  \left[
    \Omega_{\rm c}\delta_{\rm c} + (2+R)\Omega_{\rm r}\delta_{\rm r}
  \right]
   =  4 \pi G \Theta_+,
  \label{delta-two}
\end{eqnarray}
where $R=3\rho_{\rm B}/4\rho_{\rm r}$, $\Theta_+=\Theta_{00}+\Theta_{ii}$,
$a$ is the scale factor, a dot represents the derivative with respect to the conformal time $\eta$, and
the subscripts `c', `B' and `r' denote cold dark matter (CDM),
baryons, and radiation respectively.\footnote{Here we have ignored neutrinos,
whose effects on the current study is negligible.}
By splitting the perturbations into 
the initial (I) and subsequent (S) parts as
$  \delta_N ({\bf x}, \eta) = \delta_N^{\rm I}({\bf x}, \eta)
  + \delta_N^{\rm S}({\bf x}, \eta)$
where $N={\rm c, r}$, and 
employing the zero entropy fluctuation condition on super-horizon scales
as part of the initial condition,
we solve the above equations for $\Lambda=0$
in the Fourier space to yield
\begin{eqnarray}
  \widetilde{\delta}^{\rm I}_{N}({\bf k},\eta)
   & = & 
    \widetilde{\cal G}^{N}_3 (k;\eta,\eta_{\rm i})
    \widetilde{\delta}_{\rm c}({\bf k},\eta_{\rm i})
    +
    \widetilde{\cal G}^{N}_4 (k;\eta,\eta_{\rm i})
    \dot{\widetilde{\delta}}_{\rm c}({\bf k},\eta_{\rm i}),  
  \label{delta_I_N} \\ 
  \widetilde{\delta}^{\rm S}_{N}({\bf k},\eta)
  & = &
  4 \pi G \int_{\eta_{\rm i}}^{\eta} 
  \widetilde{\cal G}^{N}_4 (k;\eta,\hat{\eta})
  \widetilde\Theta_{+}({\bf k},\hat{\eta})
  \, d\hat{\eta}\,,
  \label{delta_S_N}
\end{eqnarray}
where
$\eta_{\rm i}$ is the initial conformal time, and
 the full expressions of $\widetilde{\cal G}^{N}_i$,
including the baryonic effects, are presented in \cite{cmp}.
We notice that to solve for ${\delta}_{\rm c}$ we need only two Green functions 
($\widetilde{\cal G}^{\rm c}_3$ and $\widetilde{\cal G}^{\rm c}_4$),
instead of five as presented in \cite{VeeSte}, some of which are
incorrect due to incorrect initial conditions.
Figure~\ref{fig-1} (left) shows the asymptotic behaviors of these two
Green functions on the super-horizon ($k\hat{\eta}\ll 1$) 
and sub-horizon ($k\hat{\eta}\gg 1$) scales:
$  \widetilde{T}^{\rm c}_i(k; \hat{\eta})
  \equiv \lim_{\eta/\eta_{\rm eq} \rightarrow \infty}
  \widetilde{\cal G}^{\rm c}_i(k; \eta, \hat{\eta}) {a_{\rm eq}}/{a}
$,
where the subscript `eq' denotes the epoch of radiation-matter density equality.
A simple and accurate extrapolation scheme can then be used to obtain solutions
in the non-flat or $\Lambda\neq 0$ cosmologies \cite{cmp}.
All the solutions are numerically verified to high accuracy.


One important problem for structure formation with causal seeds
is to investigate
how the source energy was compensated into the radiation and matter background
when the seeds were formed at $\eta_{\rm i}$, and
how the resulting perturbations today depend on this.
First consider the pseudo energy
$  \tau_{00}=\Theta_{00}
  +({3}/{8\pi G})\left({\dot{a}}/{a}\right)^2
  (\Omega_{\rm c}\delta_{\rm c}+\Omega_{\rm r}\delta_{\rm r})
  +({\dot{a}}/{a})\dot{\delta_{\rm c}}/4\pi G
$.
Since causality requires 
${\tau}_{00}=3{\delta}_{\rm r}/4-{\delta}_{\rm c}=0$
on super-horizon scales,
it follows that the initial source energy $\Theta_{00}$
can be compensated into
between $\delta_N$ and $\dot{\delta}_N$ with different portions.
With our Green-function solutions,
it can be straightforwardly shown that
no matter how $\Theta_{00}$
was compensated into the background at $\eta_{\rm i}$,
the resulting ${\delta_{\rm c}}^{\rm I}$ 
and thus ${\delta_{\rm c}}$ today will be the same.
This was first numerically observed in \cite{turok},
and here we can provide an analytic proof.


Finally we use our Green-function solutions to study
the compensation mechanism and the scale on which it operates.
First it can be shown that \cite{cmp}
$  \widetilde{\tau}_{00}({\bf k},\eta_0)
  = 
  \left(1-T(k)\right)
  {\widetilde\Theta_{00}({\bf k}, \eta_0)}
  +
  \int_{\eta_{\rm i}}^{\eta_0}
  \left[
      {T'}(k; \hat{\eta})
      ({\dot{a}(\hat{\eta})}/{a(\hat{\eta})})
      \widetilde\Theta_+({\bf k}, \hat{\eta})
      +
      T(k)
      \dot{\widetilde\Theta}_{00}({\bf k}, \hat{\eta})     
  \right]
  d\hat{\eta}
$,
where
$T(k)$ is the standard CDM transfer function in the inflationary models, and
$  {T'}(k; \hat{\eta})
  =
  {\widetilde{\cal G}^{\rm c}_4(k; \eta_0, \hat\eta)}/
  {\widetilde{\cal G}^{\rm c}_{40}(k; \eta_0, \hat\eta)}
$
(see Figure~\ref{fig-1} right).
On super-horizon scales,
it is clear that
$T(k)$ is unity by definition so that only the integral survives,
and 
that the quantity inside the square brackets is nothing but 
$\widetilde\Theta_{0i,i}({\bf k}, \hat{\eta})$ 
due to the source stress-energy conservation
and the fact $T=T'$.
Since $\widetilde\Theta_{0i,i}$ has a $k^4$ fall-off power spectrum 
due to causality, it follows immediately that
the pseudo-energy  $\tau_{00}$ 
also has a $k^4$-decay power spectrum outside the horizon.
On sub-horizon scales,
on the other hand,
although $(1-T(k))$ is approximately unity,
the usual sub-horizon power-law decay in 
$\widetilde\Theta_{00}$ 
(as in the case of topological defects)
will make the first term negligible,
while the quantity inside the integral is no longer simply
$\widetilde\Theta_{0i,i}$.
As a result,
we see that the compensation scale,
above which the power of both $\delta_{\rm c}$ and $\tau_{00}$ decays as $k^4$,
 is determined 
not only by the functions $T$ and $T'$, 
but also by the properties of the source near the horizon scale.
Once the detailed behavior of the source near the horizon scale is known,
we can accurately locate the compensation scale 
using our formalism \cite{cmp}.
We acknowledge the support from 
NSF KDI Grant (9872979) and
NASA LTSA Grant (NAG5-6552).



\begin{references}

\bibitem{Guth}
  Guth, A.\ H., {\it Phys.Rev.}, {\bf D23}, {347} (1981).

\bibitem{VilShe}
  For a review see Vilenkin, A.,  Shellard, E.\ P.\ S., {\it Cosmic strings and other topological defects}, Cambridge University Press, Cambridge, 1994.

\bibitem{maxiboom}
  Jaffe, A.\ H.\ et al., {\it Phys.Rev.Lett.} in press,
  {\it astro-ph/0007333} (2001).

\bibitem{AveShe4}
  Avelino, P.\ P., Shellard, E.\ P.\ S., Wu, J.\ H.\ P., Allen, B., {\it ApJ.},
{\bf 507}, {L101} (1998). 

\bibitem{HuSelWhi}
  Hu, W., Seljak, U., White, M.,  Zaldarriaga, M., {\it Phys.Rev.}, {\bf D57}, {3290} (1998). 

\bibitem{HuWhi}
  Hu, W.,  White, M., {\it Phys.Rev.}, {\bf D56}, {596} (1997).

\bibitem{cmp}
  Wu, J.\ H.\ P., {\it astro-ph/0012205} (2000).

\bibitem{VeeSte}
  Veeraraghavan, S., Stebbins, A., {\it ApJ.}, {\bf 365}, {37} (1990).

\bibitem{turok}
  Pen, U., Spergel, D., Turok, N., {\it Phys.Rev.}, {\bf D49}, {692} (1994).

\end{references}
\end{document}